\documentclass[aps,prl,preprint,showpacs,groupedaddress]{revtex4-1}

\usepackage[intlimits]{amsmath}
\usepackage[dvips]{graphicx}
\usepackage{amsfonts}
\usepackage{amssymb}
\usepackage{graphicx}
\usepackage{bm}
\usepackage{natbib}

\begin{document}

\title{Theoretical Analysis of Dynamic Processes for Interacting Molecular Motors}

\author{Hamid Teimouri $^{1}$}
\author{Anatoly B. Kolomeisky $^{1}$}
\author{Kareem Mehrabiani $^{1}$,$^{2}$}

\email{tolya@rice.edu}
\affiliation{$^1$Department of Chemistry and Center for Theoretical Biological Physics, Rice University, Houston, Texas, 77005, USA}
\affiliation{$^2$Department of Physics and Astronomy, California State University, Northridge,  California, 91330, USA}

\date{\today}

\begin{abstract}

Biological transport is supported by collective dynamics of enzymatic molecules that are called motor proteins or molecular motors. Experiments suggest that motor proteins interact locally via short-range potentials.  We investigate the fundamental role of these interactions by analyzing a new class of totally asymmetric exclusion processes where interactions are accounted for in a thermodynamically consistent fashion. Theoretical analysis that combines various mean-field calculations and computer simulations suggests that dynamic properties of molecular motors strongly depend on interactions, and correlations are stronger for interacting motor proteins. Surprisingly, it is found that there is an optimal strength of interactions (weak repulsion) that leads to a maximal particle flux. It is also argued that molecular motors transport is more sensitive to attractive interactions. Applications of these results for kinesin motor proteins are discussed.

\end{abstract}

\maketitle

A central part in supporting many cellular processes is played by several classes of enzymatic molecules that are known as motor proteins or molecular motors \cite{alberts_book,howard_book,kolomeisky07,chowdhury13,kolomeisky13}. They use the chemical energy  released from hydrolysis of adenosine triphosphate (ATP) to drive cellular transport along cytoskeleton filaments. Single-molecule properties of various molecular motors are now well investigated both experimentally and theoretically \cite{veigel11,chowdhury13,kolomeisky13}. However, cellular cargoes are often moved by groups of motor proteins, and microscopic mechanisms of collective motor behaviors remain not well understood \cite{Uppulury12,kolomeisky13,Neri13}.

Recent experiments on kinesin motor proteins indicate that  motors bound to the microtubule filament interact with each  other \cite{Roos08,Vilfan01,Seitz06}. The evidences for this behavior are found from observations that kinesins on microtubules phase segregate into more dense and less dense patches, and from measurements of different times to be bound to the filament depending on the presence of neighbors \cite{Roos08,Vilfan01,Seitz06}. It was estimated that these interactions are weakly attractive ($1.6\pm0.5 k_{B}T$) \cite{Roos08}.  It raises a question on fundamental role of this phenomenon in collective motion of motor proteins. Various chemical transitions such as bindings, unbindings, hydrolysis and steppings should be affected by this potentials, influencing the overall dynamics of motor proteins.  However, the impact of such interactions on  transport of molecular motors is not fully explored \cite{Uppulury12}. There are several investigations addressing collective dynamics of interacting motor proteins \cite{Campas06,Pinkoviezky13}. But the main limitation of these studies is a phenomenological description of interactions that does not provide a comprehensive picture for all chemical transitions in motor proteins. 

One of the most powerful tools in investigating multi-particle non-equilibrium systems is a class of models called totally asymmetric simple exclusion processes (TASEP) \cite{Derrida98,Chou11,Bressloff13}. It is known that these models successfully capture essential properties of a large number of physical, chemical and biological systems \cite{Chou11,Bressloff13,Frey03,Dong12,Golubeva12,tsekouras08,klumpp03}. Different versions of TASEP have been extensively employed in studies of various aspects of biological molecular motors \cite{Chou11,Pinkoviezky13,Neri13,Dong12}, providing an important microscopic insights on these complex processes. TASEP with interactions have been studied before, but only for the particles on the ring \cite{Pinkoviezky13} or with phenomenologically defined interactions \cite{Antal00}.

In this letter, we investigate the effect of intermolecular interactions on collective dynamics of motor proteins by introducing a new TASEP model that treats interactions in a thermodynamically consistent way. To make the model more realistic we use open boundary conditions since the cytoskeleton filaments have finite length. Using various mean-field analytical methods and extensive Monte Carlo simulations we compute particle currents and density profiles for  molecular motors. It provides us with a direct method to address the fundamental role of interactions. Our analysis suggests that there is an optimal interaction strength, corresponding to weak attractions, that leads to the maximal particle flux. It is also found that interactions introduce significant correlations in the system and modify phase diagrams.  In addition, dynamic properties of molecular motors are influenced stronger by attractive interactions.  

\begin{figure}
\vspace*{-0.2cm}
\hspace*{-0.1cm}
\centering
 \includegraphics[clip,width=0.45\textwidth]{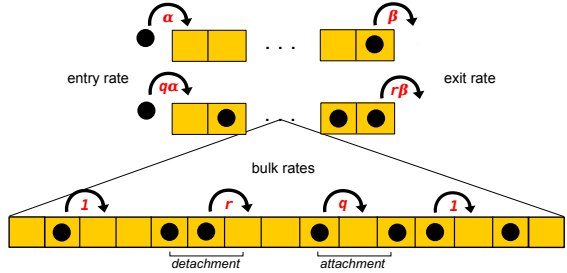}
\caption{A schematic view of TASEP model with interacting particles.}
\end{figure}

 We consider a transport of molecular motors on the cytoskeleton filaments as a multi-particle motion along a lattice segment with $L$ sites as illustrated in Fig. 1. For each lattice site $i$  ($1 \leq i \leq L$) we assign an occupation number $\tau_{i}$, which is zero if the site is empty or $\tau_{i}=1$ if the site is occupied. Each site cannot be occupied by more than one particle. It is assumed that each two particles sitting on neighboring sites interact with each other with an energy $E$ ($E>0$ correspond to attractions and $E<0$ describe repulsions). A single motor that is not a part of the particles cluster can move forward with the rate 1  if it moves to the site without neighbors (Fig. 1). There is no energy change in this case. However, if the particle hops into another cluster it moves with  rate $q \ne 1$ because the energy of the system changed by creating a new pair of neighbors (see Fig. 1). Similarly, for the particle breaking from the cluster its forward rate is equal to $r \ne 1$ when the particle does not have neighbors in the new position. But for the case when one pair is broken and another one is created the stepping rate is equal to 1 since there is no overall energy change (Fig. 1). Creating and breaking the pair of particles can be viewed as opposite chemical transitions, so  the detailed balance arguments can be applied,
\begin{equation}\label{db}
\frac{q}{r}=\exp{(\frac{E}{k_{B}T})}.
\end{equation} 
To simplify analysis, we assume that the energy $E$ is equally split between creation and breaking processes, providing explicit expressions for the stepping rates $q$ and $r$, 
\begin{equation}\label{rates}
q=\exp{(\frac{E}{2 k_{B}T})}, \quad r=\frac{1}{q}=\exp{(\frac{-E}{2k_{B}T})}.
\end{equation} 
The splitting of the interaction potential between the rates $q$ and $r$ is not unique, but other possibilities can be easily explored in our method. In addition, it can be shown that particle dynamics is similar for all cases. Eqs. (\ref{rates}) have a clear physical meaning. For attractive interactions ($E>0$) the particle moves faster ($q>1$) to create a new pair since the energy of the system decreases by $E$. Breaking out of the cluster increases the energy by $E$ and the transition rate is slower ($r<1$). Similar arguments can be given for repulsive interactions ($E <0$). When there is no interactions ($E=0$) we have $q=r=1$ and the original TASEP with only hard-core exclusions is recovered. It is important to note that, in contrast to previous studies, this is a thermodynamically consistent approach that accounts for interactions in all transitions in the system. In addition, it differs from other TASEP with interactions \cite{Antal00,Pinkoviezky13} because the stepping rates depend on the state of 4 consecutive lattice sites. Interactions also modify the boundary transitions as shown in Fig. 1. The entrance rate is equal to $\alpha$ if no particle pair created, while the rate is equal to $q \alpha$ when the pair creation is involved. Similarly, the  exit rate of the single particle is given by the rate $\beta$, while exiting with breaking from the cluster changes the rate to $r \beta$.

To analyze the system we start with the simplest mean-field  (SMF) approach that neglects all correlations in the system. It assumes that for any two sites on the lattice their occupancies are independent of each other, i.e., $Prob(\tau_{i},\tau_{j}) \approx Prob(\tau_{i})* Prob(\tau_{j})$ for $1 \le i,j \le L$. The particle density at every site is associated with an average occupancy, $\rho=<\tau>$, and it reaches a constant value in the bulk of the system. It can be shown that, similarly to the classical TASEP without interactions, there are three stationary phases, low density (LD), high density (HD) and maximal current (MC), as illustrated in Fig. 2. In the LD phase the particle bulk density and the current are equal to \cite{sm},
\begin{equation}\label{LD}
\rho_{LD}=\frac{q-\sqrt{q^{2} - 4 \alpha q (q-1)}}{2(q-1)},\quad J_{LD}=\alpha \left( \frac{\alpha q(q-1)- 1 + \sqrt{q^{2} - 4\alpha q(q-1)}}{q-1} \right).
\end{equation}
In the HD phase the calculations yield,
\begin{equation}\label{HD}
\rho_{HD}=\frac{q-2+\sqrt{q^{2}-4\beta(q-1)}}{2(q-1)}, \quad J_{HD}=\beta \left( \frac{\beta(q-1)-1+\sqrt{q^{2}-4\beta(q-1)}}{q(q-1)} \right).
\end{equation}
In the MC phase the bulk density reaches the maximal value of $\rho_{MC}=1/2$, while the particle current can be written as
\begin{equation}\label{MC}
J_{MC}=\frac{1}{8}+ \frac{r+q}{16}.
\end{equation}
As expected, for $E=0$ ($q=r=1$) these equations yield the results for the standard TASEP with only hard-core exclusions. Comparing theoretical predictions of SMF approach with Monte Carlo computer simulations (Figs. 2 and 3) we can see that  it is a reasonable approximation for very weak interactions ($E \approx 0$), while for stronger attractions or repulsions the simple mean-field method does not work well. The calculated in SMF density profiles deviate from computer simulations results (see Fig. 3). But the strongest argument against using SMF for TASEP with interactions comes  from the analysis of Eq. (\ref{MC}) for the current in the MC phase. It predicts that for $|E| \gg 1$ the current is increasing without a bound, which is clearly an unphysical result. In the case of strong attractions particles will tend to stay in one big cluster that cannot move because particle breaking from the cluster is not possible. In this case the current is expected to go to zero. For strong repulsions no particle pairs can be created and the flux is rate limited by the exit process from the last lattice site, i.e., $J_{MC} \sim 1/L \rightarrow 0$, in the thermodynamic limit of $L \gg 1$.

The fact that SMF method neglects correlations is the main reason for not satisfactory description of TASEP with stronger intermolecular interactions. To develop a more reasonable analysis we propose to use a mean-field approach that takes into account some correlations. Our idea is to fully describe particle dynamics inside of a cluster of several lattice sites, but correlations between states of different clusters will be neglected. In our calculations clusters with 2 lattice sites are utilized. In this approach, the occupation of four consecutive sites is written as  $Prob(\tau_{i-1},\tau_{i}, \tau_{i+1},\tau_{i+2}) \approx Prob(\tau_{i-1},\tau_{i})*Prob(\tau_{i+1},\tau_{i+2})$.  The method is called a cluster mean-field (CMF). There are four possible states for each two-site cluster depending on the occupancy of sites that can be labeled as (1,1), (1,0), (0,1) and (0,0). We define $P_{11}$, $P_{10}$, $P_{01}$ and  $P_{00}$ as probabilities to be found in one of these configurations, respectively. The normalization requires that $P_{11}+P_{10}+P_{01}+P_{00}=1$. The average bulk density and the current can be expressed in terms of these functions \cite{sm},
\begin{equation}
\rho_{bulk}=P_{11}+\frac{P_{01}+P_{10}}{2}; \quad J=q P_{01} P_{01}+ r P_{11} P_{00} + P_{11} P_{01}+P_{01} P_{00}.
\end{equation}   
In CMF all dynamic properties for TASEP with interactions can be obtained from the temporal evolution of cluster probabilities, as presented in the Supplemental Material \cite{sm}. More specifically, for (1,1) cluster in the bulk the master equation can be  written as
\begin{equation}
\frac{dP_{11}}{dt}=q P_{01}P_{01}+P_{11}P_{01} - rP_{11}P_{00} -P_{11}P_{01}.
\end{equation}
At the entrance the dynamics of (1,1) clusters follows
 \begin{equation}
\frac{dP_{11}}{dt}=q \alpha P_{01} -r P_{11}P_{00}-P_{11}P_{01},
\end{equation}
while at the exit we have 
\begin{equation}
 \frac{dP_{11}}{dt}=q P_{01} P_{01}+P_{11} P_{01} - r \beta P_{11}.
\end{equation}
Similar expressions can be written for clusters (1,0), (0,1) and (0,0). At large times these master equations can be solved numerically exactly, from which all dynamic properties can be estimated. The results for various properties are shown in Figs. 2-4. Our analysis again finds three stationary phases (see Figs. 2 and 3). One can see that the CMF method provides a much better agreement with predictions from computer simulations for all dynamic properties.

\begin{figure}
\vspace*{-0.2cm}
\hspace*{-0.1cm}
\centering
 \includegraphics[clip,width=0.45\textwidth]{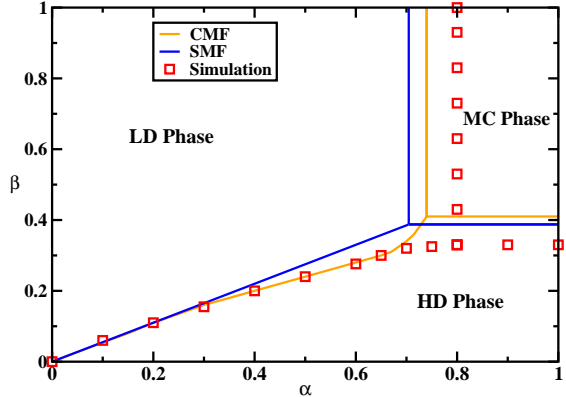}
\caption{Stationary phase diagram for TASEP with intermolecular interactions. The case of weakly repulsive interactions, $E=-1.2$ $k_{B}T$, is shown.}
\end{figure}

Theoretical framework of the CMF method along with computer simulations  allows us to investigate the fundamental effect of interactions on multi-particle dynamics in the TASEP model. It has been argued above that particle currents should diminish for strong attractions and repulsions. It suggests that there is an intermediate strength of interactions where the maximal flux might be achieved. Our calculations support these arguments as illustrated in Fig. 4. We found that this optimal strength  corresponds to weak repulsions with $E^{*} \approx -3$ $k_{B}T$ in CMF, while the simulations indicate  $E^{*} \approx -1.2$ $k_{B}T$. The surprising observation is that optimal conditions do not correspond to the case of no interactions, as one would expected from naive symmetry arguments. In addition, the optimal particle flux can be larger than the current for the system with only hard-core exclusions. The computer simulations predict $J_{max} \approx 0.29$, which is 16$\%$ more than the maximal current  for TASEP without interactions $J_{max}=0.25$.  Thus, intermolecular interactions might significantly modify particle fluxes.

\begin{figure}
\vspace*{-0.2cm}
\hspace*{-0.1cm}
\centering
 \includegraphics[clip,width=0.45\textwidth]{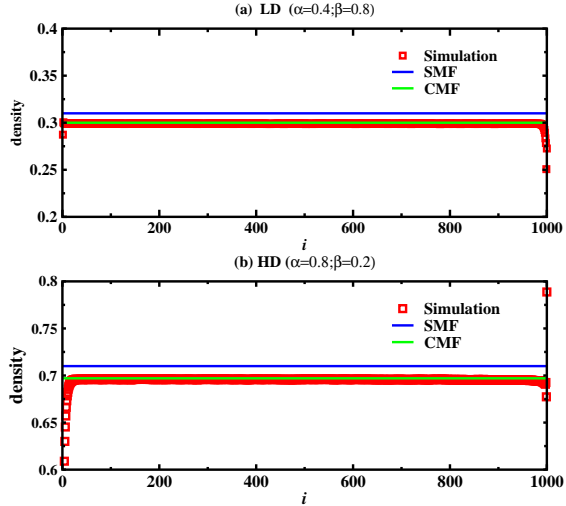}
\caption{Density profiles TASEP with intermolecular interactions for $E=-1.2$ $k_{B}T$. (a) LD phase with $\alpha=0.4$ and $\beta=0.8$  ; (b) HD phase with $\alpha=0.8$ and $\beta=0.2$}
\end{figure}

It could be also observed that the effect of interactions on particle dynamics in TASEP is not symmetric with respect to $E=0$. The results of the CMF calculations and Monte Carlo computer simulations suggest that there is more sensitivity for attractive interactions. The phase diagram also depends on the sign and strength of interactions. Fig. 5 shows the position of the triple point (that connects LD, HD and MC phases) at different values of $E$. One can see that increasing repulsions shrinks the MC and HD phases, and the LD phase occupies the largest fraction of the parameters space. For strong attractions the result is opposite. The HD phase dominates, while the LD and MC phase significantly diminish. These observations can be easily explained. Repulsions decrease the effective entrance rate into the system, making it a rate-limiting step for a larger range of parameters. This corresponds to expanding the LD phase. For attractions the exit rate slows down significantly because particles leaving the system sometimes should break from the clusters. This is not favorable from the energetic point of view. In this case, the HD phase dominates the system.

\begin{figure}
\vspace*{-0.2cm}
\hspace*{-0.1cm}
\centering
 \includegraphics[clip,width=0.45\textwidth]{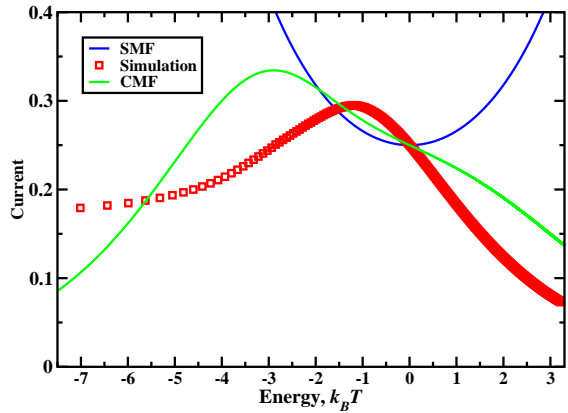}
\caption{Maximal current as a function of the interaction strength. Lines are predictions from mean-field  calculations. Symbols are from Monte Carlo computer simulations.}
\end{figure}

It is interesting to apply our theoretical analysis for real motor proteins. Experimental studies show that kinesins molecular motors bound to cytoskeleton filaments experience a short-range attractive interactions of order  $E=1.6\pm 0.5 k_{B}T$ \cite{Roos08}. Comparing this with plots in Fig. 4 we conclude that kinesins probably do not function at the most optimal regime with the maximal particle current. However, they operate at conditions where small changes in interactions might lead to large modifications in dynamic properties. It suggests that kinesins might be optimized not for the maximal flux but for supporting robust cellular transport via tuning its intermolecular interactions. It allows molecular motors to compensate for fluctuations due to collisions with other molecules and from external loads.  This picture agrees with current views on mechanisms of cooperativity in multiple kinesins \cite{kolomeisky13,Uppulury12}. However, we should notice that our model of motor protein dynamics is oversimplified. It ignores many important processes such as back steppings, bindings to the filaments and unbindings from them, and hydrolysis. It is not clear what effect the intermolecular interactions will have if all relevant chemical transitions are included.

\begin{figure}
\vspace*{-0.2cm}
\hspace*{-0.1cm}
\centering
 \includegraphics[clip,width=0.45\textwidth]{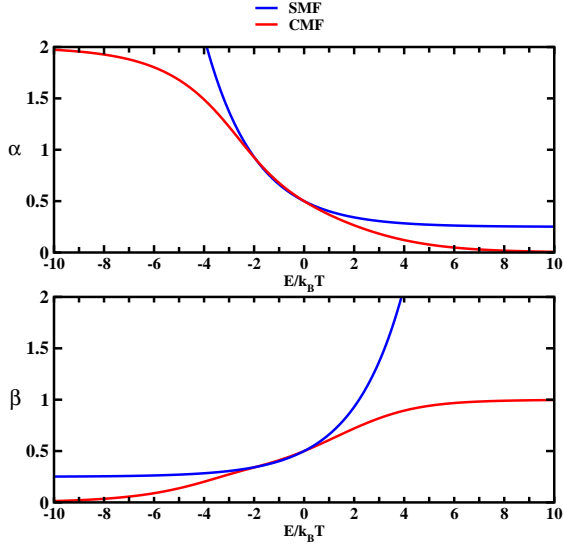}
\caption{Coordinates of the triple point as a function of interaction strength. Lines correspond to predictions from mean-field calculations. Computer simulations results are close to CMF predictions.}
\end{figure}

In conclusion, we developed a new theoretical approach to investigate the effect of intermolecular interactions on dynamics of cellular molecular motors that move along cytoskeleton filaments. Our method is based on employing totally asymmetric simple exclusion processes that are known to be successful for analysis of non-equilibrium multi-particle phenomena. The important part of the method is a thermodynamically consistent procedure that allowed us to quantitatively describe  the effect of intermolecular interactions. Theoretical calculations indicate that interactions bring significant spatial correlations in the system that could be partially captured by considering dynamics of clusters. It is found that there is an optimal strength of interactions at which the particle current reaches the maximum, while for large attractions or repulsions the fluxes disappear. For TASEP these optimal conditions correspond to weak repulsions. This observation is unexpected since from naive symmetry arguments the case of no interactions seems to be optimal. Interactions also modify stationary phase diagrams. For repulsions the LD phase becomes the most important, while for attractions the HD phase dominates. Our analysis also show that dynamic properties are more sensitive to attractive interactions. The implications of these observations for kinesins motor proteins are discussed. It is argued that kinesins might be functioning under conditions to support the robustness of the cellular transport instead of the maximal fluxes. At the same time, it was noticed that our theoretical analysis does not account for several important transitions in motor proteins that might limit its applicability in the current form. It will be important to extend our method to include these features and to test our theoretical predictions for other classes of motor proteins.

We acknowledge support from the National Institute of Health  (grant 1R01GM094489-01) and from the Welch Foundation (grant C-1559).

\end{document}